\documentclass{llncs}
\usepackage{llncsdoc}
\usepackage{graphicx}
\usepackage{multirow}
\graphicspath{ {figures/} }
\begin{document}

\title{\mbox{OpenCL} + \mbox{OpenSHMEM} Hybrid Programming Model for the Adapteva Epiphany Architecture}

\author{
	David A. Richie\inst{1}
	\and
	James A. Ross\inst{2}
}

\institute{
	Brown Deer Technology, Forest Hill, MD 21050, USA
	\email{drichie@browndeertechnology.com}
	\and
	U.S. Army Research Laboratory, Aberdeen Proving Ground, MD 21005, USA
	\email{james.a.ross176.civ@mail.mil}
}

\maketitle

\begin{abstract}
There is interest in exploring hybrid \mbox{OpenSHMEM} + X programming models
to extend the applicability of the \mbox{OpenSHMEM} interface to more hardware
architectures. We present a hybrid \mbox{OpenCL} + \mbox{OpenSHMEM} programming
model for device-level programming for architectures like the Adapteva Epiphany
many-core RISC array processor. The Epiphany architecture comprises a 2D array
of low-power RISC cores with minimal uncore functionality connected by a 2D
mesh Network-on-Chip (NoC). The Epiphany architecture offers high computational
energy efficiency for integer and floating point calculations as well as
parallel scalability. The Epiphany-III is available as a coprocessor in
platforms that also utilize an ARM CPU host. \mbox{OpenCL} provides good
functionality for supporting a co-design programming model in which the host
CPU offloads parallel work to a coprocessor. However, the \mbox{OpenCL} memory
model is inconsistent with the Epiphany memory architecture and lacks support
for inter-core communication. We propose a hybrid programming model in which
\mbox{OpenSHMEM} provides a better solution by replacing the non-standard
\mbox{OpenCL} extensions introduced to achieve high performance with the
Epiphany architecture. We demonstrate the proposed programming model for
matrix-matrix multiplication based on Cannon's algorithm showing that the
hybrid model addresses the deficiencies of using \mbox{OpenCL} alone to achieve
good benchmark performance.

\begin{keywords}
OpenCL, OpenSHMEM, hybrid programming model, single-board computer,
Network-on-Chip (NoC)
\end{keywords}

\end{abstract}

\section{Introduction and Motivation}
\label{sec:intro}

The emergence of a wide range of parallel processor architectures continues to
present the challenge of identifying an effective programming model that
provides access to the capabilities of the architecture while simultaneously
providing the programmer with familiar, if not standardized, semantics and
syntax. The programmer is often left with the choice of using a non-standard
programming model specific to the architecture or a standardized programming
model that yields poor control and performance. The parallel RISC processor
investigated in this work has presented precisely this challenge as suitable
programming models matched to the architecture have been explored.

The Adapteva Epiphany RISC array architecture \cite{adapteva} is a scalable 2D
array of low-power RISC cores with minimal uncore functionality supported by an
on-chip 2D mesh Network-on-Chip (NoC) for fast inter-core communication. The
Epiphany architecture is scalable to 4,096 cores and represents an example of
an architecture designed for power-efficiency at extreme on-chip core counts.
Processors based on this architecture exhibit good performance/power metrics
\cite{kickstarting} and scalability via the 2D mesh network \cite{wentzlaff}
\cite{taylor}, but require a suitable programming model to fully exploit these
capabilities. A 16-core Epiphany-III coprocessor~\cite{datasheet} has been
integrated into the Parallella mini-computer platform \cite{reference} where
the RISC array is supported by a dual-core ARM CPU and asymmetric shared-memory
access to off-chip global memory.

RISC array processors such as those based on the Epiphany architecture may
offer significant computational power efficiency in the near future with
requirements in increased floating point performance, including long-term plans
for exascale platforms. The power efficiency of the Epiphany architecture has
been specifically identified as both a guide and prospective architecture for
such platforms \cite{varghese}. The Epiphany-IV processor has a performance
efficiency of 50 GFLOPS/W (single precision) \cite{kickstarting} making it one
of the most efficient fully divergent parallel processors based on
general-purpose cores. This approaches the threshold for exascale computing
requirements of a power budget of 20 megawatts \cite{bergman}. This
architecture has characteristics consistent with future processor predictions
arguing hundreds \cite{fewcores} and thousands \cite{asanovic}, \cite{borkar}
of cores on a chip.

One aspect of the low-power design of the Epiphany architecture is the use of a
cache-less distributed on-chip memory architecture that for the Epiphany-III
provides 32 KB of local memory per core for both instructions and data.
Utilizing this core local memory and managing inter-core communication is
critical to achieving good performance and this is a central element in the
design of the architecture. In previous work, these technical challenges were
the primary factors in achieving good performance with threaded MPI and less
favorable results using \mbox{OpenCL}. Here we revisit \mbox{OpenCL} with a
hybrid model that uses \mbox{OpenSHMEM} to resolve the deficiencies of
\mbox{OpenCL} in the context of this architecture. Our main contributions are
the presentation of a hybrid \mbox{OpenCL} + \mbox{OpenSHMEM} programming model
with benchmarks for the application to matrix-matrix multiplication.

An outline of the remainder of the paper is as follows. Section
\ref{sec:background} describes the Epiphany architecture and previous work
using \mbox{OpenCL} and \mbox{OpenSHMEM} as parallel programming models.
Section \ref{sec:hybrid} presents the proposed hybrid \mbox{OpenCL} +
\mbox{OpenSHMEM} programming model for device-level programming. Section
\ref{sec:application} discusses the application of the proposed programming
model to Cannon's algorithm for matrix-matrix multiplication, including
benchmark results. Section \ref{sec:conclusions} discusses conclusions and
future work.

\section{Background}
\label{sec:background}

Interest in exploring hybrid \mbox{OpenSHMEM} + X programming models has been
expressed recently within the \mbox{OpenSHMEM} community \cite{Baker}. Just as
the two-tier parallel hybrid OpenMP + MPI model handles both symmetric
multiprocessing (SMP) execution within a node and distributed message passing
for attached network nodes, it is assumed that similar hybrid models may
benefit from mixing code with \mbox{OpenSHMEM}. In the specific case detailed
within this paper, the hybrid \mbox{OpenCL} + \mbox{OpenSHMEM} model exists at
the same parallelism tier and the combination of the programming models address
the deficiencies of each within the context of the Parallella platform and
Epiphany architecture.  While \mbox{OpenCL} may do well addressing SMP
architectures with hierarchical memory, it does not provide semantics for
inter-processor communication between processing elements or multiprocessors.
\mbox{OpenSHMEM} provides the semantics for non-uniform memory access (NUMA)
across a partitioned global address space (PGAS) and may not be ideal for SMP
architectures. The \mbox{OpenSHMEM} concept of memory exists virtually in a
flat one-dimensional domain and lacks the semantics of the tiered memory
hierarchy found in the SMP-based \mbox{OpenCL} programming model.
Fundamentally, the Epiphany device-level architecture has characteristics of
both SMP and PGAS platforms so it makes sense to address the architecture with
a hybrid SMP and PGAS programming model.

\subsection{Epiphany Architecture}
\label{ssec:epiphany}

\begin{figure}
	\centering
		\includegraphics[width=1.0\textwidth]{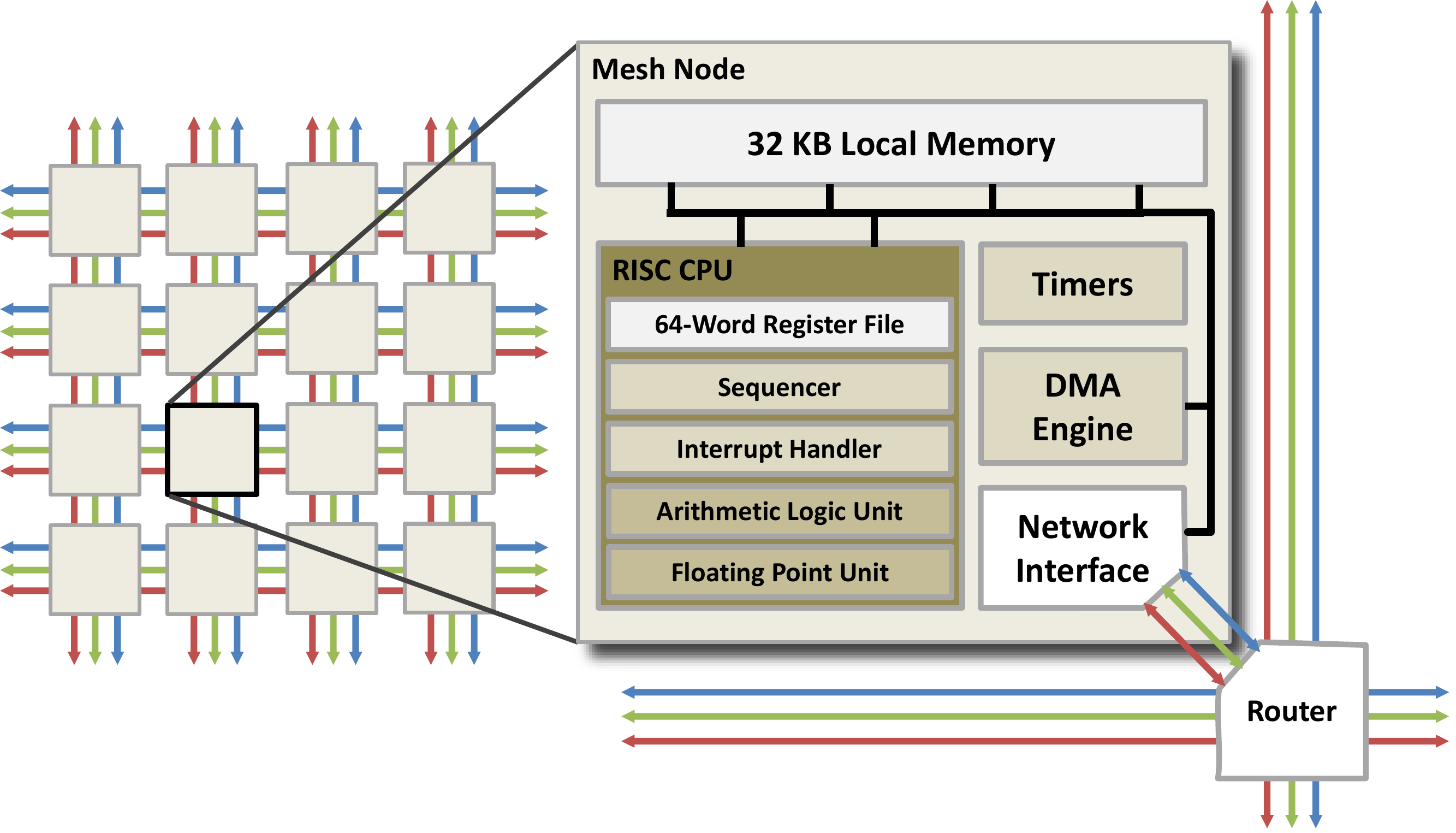}
	\caption{Adapteva Epiphany-III architecture diagram}
	\label{fig:epiphany}
\end{figure}

The Adapteva Epiphany MIMD architecture is a scalable 2D array of RISC cores
with minimal uncore functionality connected with a fast 2D mesh Network-on-Chip
(NoC). Processors based on this architecture exhibit good energy efficiency and
scalability via the 2D mesh network, but require a suitable programming model
to fully exploit the architecture. The 16-core Epiphany-III coprocessor has
been integrated into the Parallella minicomputer platform where the RISC array
is supported by a dual-core ARM CPU and asymmetric shared-memory access to
off-chip global memory. Figure \ref{fig:epiphany} shows the high-level
architectural features of the coprocessor. Each of the 16~Epiphany-III mesh
nodes contains 32~KB of shared local memory (used for both program instructions
and data), a mesh network interface, a dual-channel DMA engine, and a RISC CPU
core. Each RISC CPU core contains a 64-word register file, sequencer, interrupt
handler, arithmetic logic unit, and a floating point unit. Each processor tile
is very small at 0.5~mm\textsuperscript{2} on the 65~nm process and
0.128~mm\textsuperscript{2} on the 28~nm process. Peak single-precision
performance for the Epiphany-III is 19.2~GFLOPS with a 600~MHz clock.
Fabricated on the 65~nm process, the Epiphany-III consumes 594~mW for an energy
efficiency of 32.3~GFLOPS per watt [Olofsson, personal communication]. The
64-core Epiphany IV, fabricated on the 28~nm process, has demonstrated energy
efficiency exceeding 50~GFLOPS per watt \cite{kickstarting}.

The raw performance of currently available Epiphany coprocessors is relatively
low compared to modern high-performance CPUs and GPUs; however, the Epiphany
architecture provides greater energy efficiency and is designed to be highly
scalable. The published architecture road map specifies a scale-out of the
architecture to exceed 1,000 cores in the near future and, shortly thereafter,
tens of thousands of cores with an energy efficiency approaching one TFLOPS per
watt. Within this context of a highly scalable architecture with high energy
efficiency, we view it as a competitive processor technology comparable to GPUs
and other coprocessors.

While architecture energy efficiency is important, achievable performance with
a compelling programming model is equally, if not more, important. Key to
performance with the Epiphany architecture is data re-use, requiring precise
control of inter-core communication since the architecture does not provide a
hardware cache at any level. The cores can access off-chip mapped memory with a
significant performance penalty in both latency and bandwidth relative to
accessing on-chip core memory of any core.

\subsection{\mbox{OpenCL} for Epiphany}
\label{ssec:opencl}

\mbox{OpenCL} is an industry standard API for parallel programming accelerators
or coprocessors on heterogeneous platforms \cite{stone}. Designed primarily for
computing with general-purpose graphics processing units (GPUs), the API may be
used to access the compute capability of other types of devices including
multi-core CPUs and other accelerators. \mbox{OpenCL} support is provided for
most mainstream high-performance computing accelerators including Nvidia and
AMD GPUs, Intel and AMD multi-core CPUs, Intel Xeon Phi, and mobile CPU+GPU
hybrid processors. In this context, there is merit in exploring the use of
\mbox{OpenCL} for exposing the compute capability of the Epiphany coprocessor
on the Parallella.

\mbox{OpenCL} consists of a kernel programming API used to program the
coprocessor device and a run-time host API used to coordinate the execution of
these kernels and perform other operations such as memory synchronization so
that parallel computationally intensive work can be offloaded from the host
platform. The \mbox{OpenCL} programming model is based on the parallel
execution of a kernel over many threads to exploit SIMD or SIMT architectures.
From the perspective of the host platform, parallel kernels are enqueued for
execution on the coprocessor device. Each kernel is executed over a global
n-dimensional range of work items logically partitioned into local workgroups.
Threads of execution within a workgroup are allowed limited synchronization
through the use of barriers, and no synchronization between workgroups is
allowed.

\mbox{OpenCL} was the first standard parallel programming API implemented for
the Epiphany architecture, and partial support for the \mbox{OpenCL} 1.1
standard was available as part of the COPRTHR-1.5 SDK for Epiphany
\cite{coprthr}. The selection of \mbox{OpenCL} was supported by several
factors. The Epiphany-III coprocessor was available as part of a heterogeneous
mini-computer (Parallella) that included a dual-core ARM CPU host running
Linux. As a result, the \mbox{OpenCL} co-design programming model premised on
the host-directed offload of parallel work to a coprocessor was well suited to
the platform.

The focus of the implementation of \mbox{OpenCL} for Epiphany was to leverage
the API to support effective parallel programming and take advantage of the
underlying architecture. As with other non-GPU architectures, limitations and
constraints exist in the use of \mbox{OpenCL} for targeting the Epiphany
architecture. \mbox{OpenCL} was designed for massively multithreaded
architectures such as GPUs. However Epiphany has no hardware support for
multithreading and early experiments with software supported multithreading
were not successful due in part to resource constraints. As a result,
implementation of the \mbox{OpenCL} device execution model for Epiphany must
constrain the workgroup size to the number of physical cores on the device.

\begin{figure}
	\centering
		\includegraphics[width=0.67\textwidth]{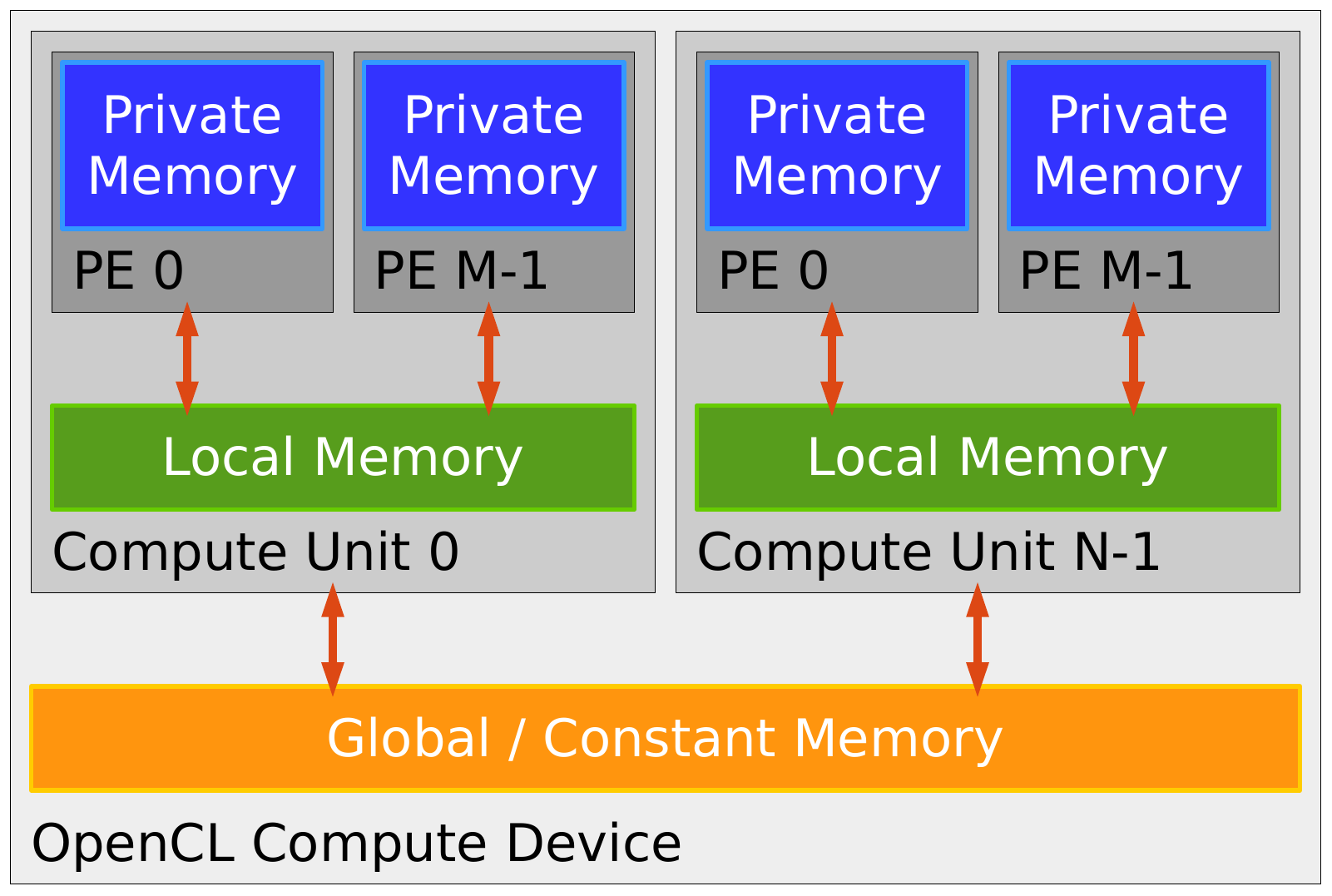}
	\caption{OpenCL memory model}
	\label{fig:opencl}
\end{figure}

The most significant technical issue encountered in the implementation of
\mbox{OpenCL} for Epiphany was reconciling the physical memory architecture of
the Epiphany coprocessor with the logical memory model defined by the
\mbox{OpenCL} standard, shown in Figure \ref{fig:opencl}. \mbox{OpenCL} address
space qualifiers co-mingle the concepts of physical locality and visibility.
For the Epiphany architecture, the physical memory co-located with each core
executing a thread in an \mbox{OpenCL} workgroup is best described as symmetric
distributed shared memory. This memory is physically local to the executing
thread while also having shared visibility with all other threads since remote
cores have non-uniform memory access to the local memory of any core. Managing
the use and re-use of this symmetric distributed memory is critical to
performance with the Epiphany architecture. An implementation treating this
memory as \mbox{OpenCL} local may prove functionally correct and consistent
within the standard, but the programmer will be left with poor performance
without an interface to treat the memory correctly. Therefore, an interface for
the symmetric distributed shared memory is needed to properly manage on-chip
data movement.

For this reason extensions were initially provided within the \mbox{OpenCL}
implementation for Epiphany. A set of inter-thread memory copy routines were
provided to allow for the direct copying of data between the local memory of
one core to another. These routines resolved the problem with \mbox{OpenCL} in
a non-standard way that nevertheless enabled algorithms to be implemented with
good performance. At the time of this development the \mbox{OpenSHMEM} standard
was close to publication but not yet released. In hindsight, \mbox{OpenSHMEM}
was precisely the interface that was needed to resolve this critical issue that
arises from the use of \mbox{OpenCL} for Epiphany.

\subsection{OpenSHMEM for Epiphany}
\label{ssec:openshmem}

\begin{figure}
	\centering
		\includegraphics[width=0.67\textwidth]{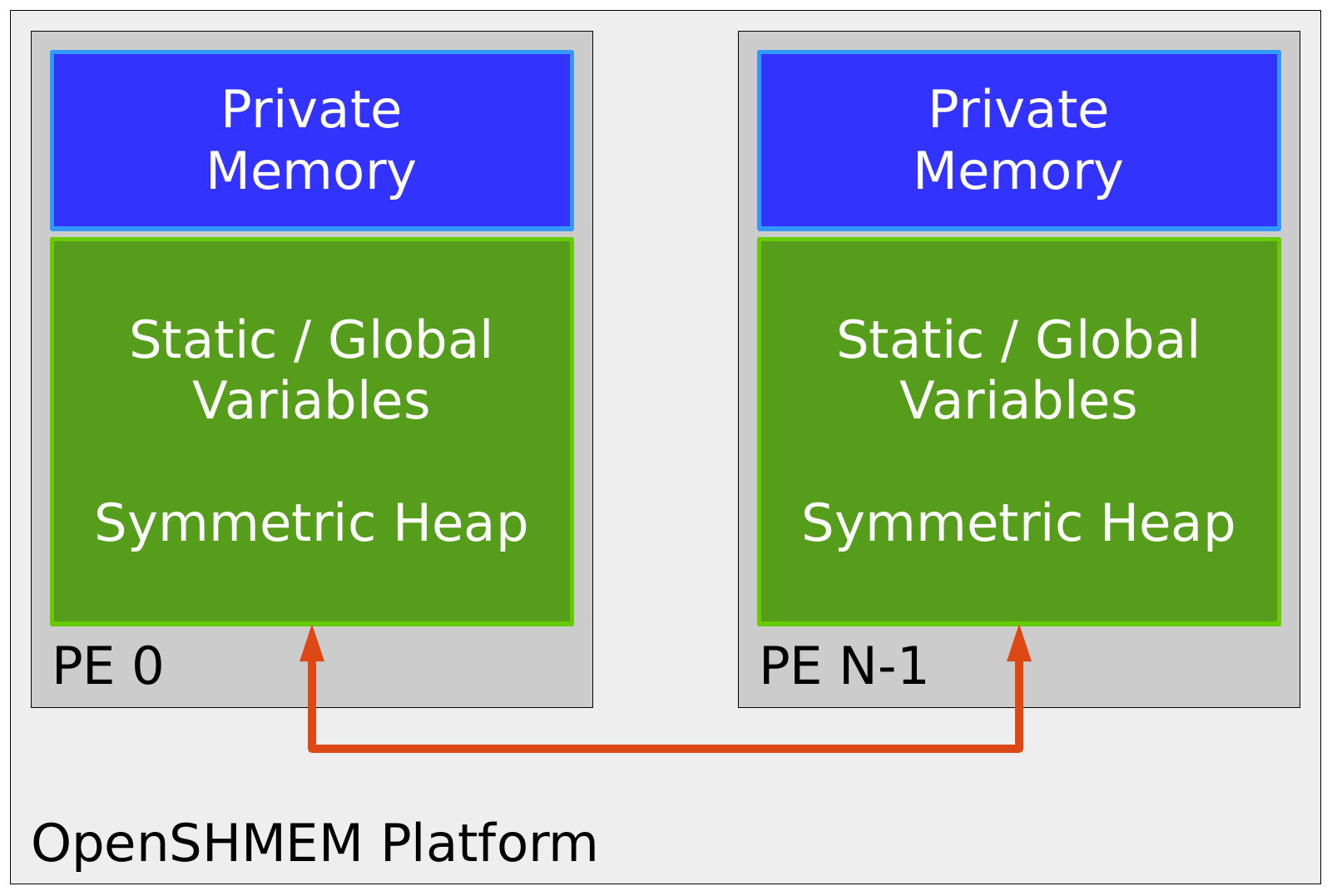}
	\caption{OpenSHMEM memory model}
	\label{fig:openshmem}
\end{figure}

An implementation of \mbox{OpenSHMEM} targeting the Epiphany architecture was
recently developed \cite{ARLOpenSHMEM}. The interface provides access the
complete \mbox{OpenSHMEM} 1.3 standard for Epiphany device-level execution. It
fills the void left by the lack of a standard programming model able to achieve
good performance with on-chip memory distributed through the NoC. Conceptually,
the physical memory of the Epiphany architecture maps directly to the
\mbox{OpenSHMEM} and PGAS memory model (shown in Figure \ref{fig:openshmem}).
The \mbox{OpenSHMEM} interface for Epiphany does not address the concept of
coprocessor offload or off-chip memory. For applications requiring these
concepts, a hybrid programming model is required.

\section{Hybrid OpenCL + OpenSHMEM Programming Model}
\label{sec:hybrid}

Based on this prior work we propose a hybrid programming model that combines
\mbox{OpenCL} with \mbox{OpenSHMEM} for device-level programming of parallel
processors like those based on the Epiphany architecture. In the simplest
terms, \mbox{OpenSHMEM} directly resolves the most critical technical issue
encountered in the implementation of \mbox{OpenCL} for such architectures, and
replaces the non-standard extensions that were originally introduced to support
inter-core data re-use and achieve good performance when implementing
algorithms for Epiphany. At the same time, \mbox{OpenCL} complements
\mbox{OpenSHMEM} in that for hybrid platforms that employ a parallel
coprocessor, \mbox{OpenCL} provides support for the \mbox{offload} of parallel
work to the coprocessor while there is no equivalent operation defined within
the \mbox{OpenSHMEM} standard.

\begin{figure}
	\centering
		\includegraphics[width=0.67\textwidth]{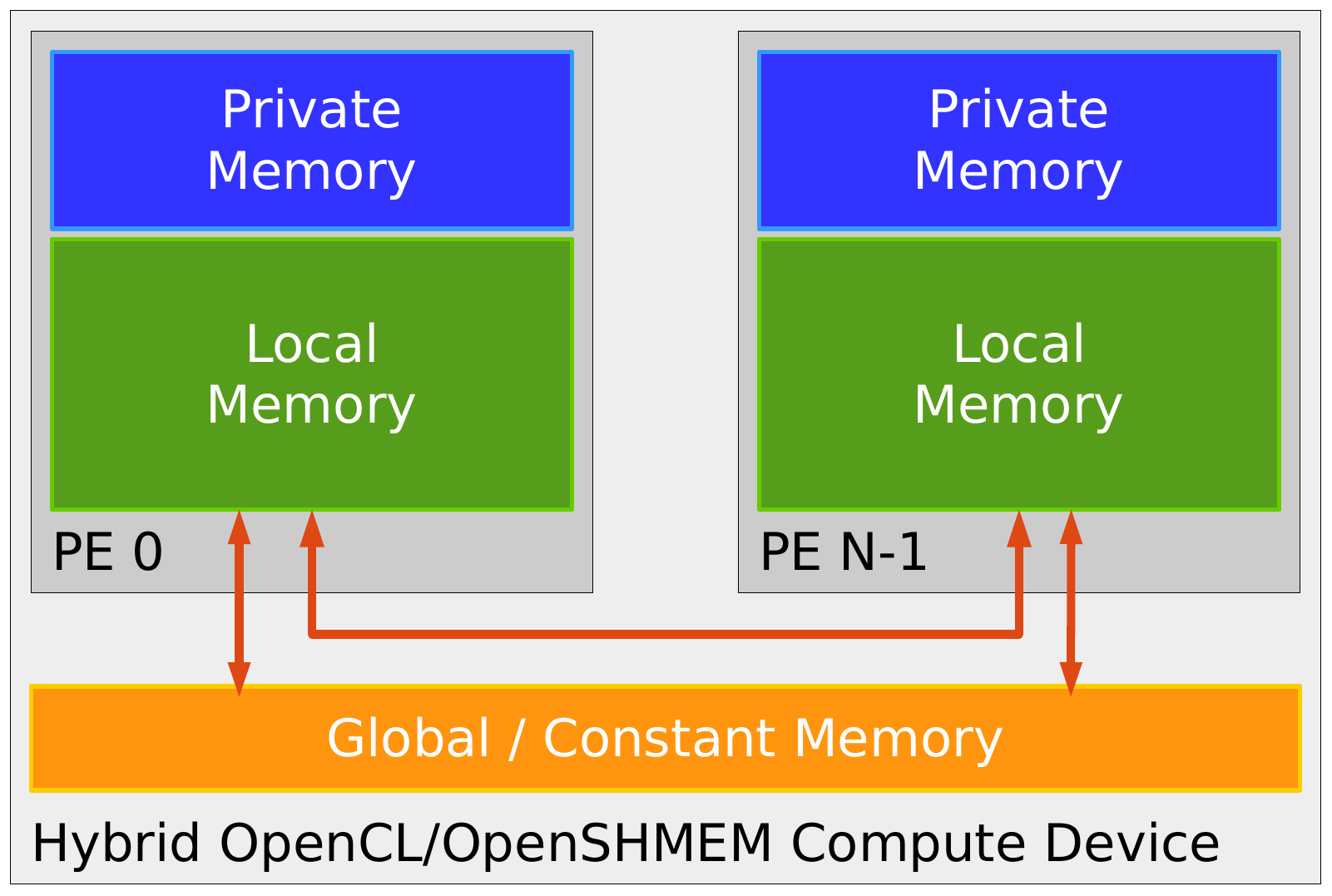}
	\caption{Hybrid OpenCL + OpenSHMEM memory model}
	\label{fig:hybrid}
\end{figure}

\mbox{OpenSHMEM} for Epiphany provides the inter-core communication between the
\mbox{OpenCL} concept of a processing element or multiprocessor. In the case of
the Epiphany architecture, they are one in the same. There is a single
processing element per multiprocessor in order to address the hierarchical
memory concept of local memory within the \mbox{OpenCL} specification. The
\mbox{OpenCL} interface defines the global or constant memory (shown in Figure
\ref{fig:hybrid})

The hybrid \mbox{OpenCL} + \mbox{OpenSHMEM} programming model uses
\mbox{OpenCL} for the development of host code that controls the overall
application and directs the operations of the coprocessor through the offload
of parallel computational kernels. The \mbox{OpenCL} kernel programming
language, closely related to standard C, is used for the implementation of
kernels. The distributed shared memory for which \mbox{OpenCL} provides no
suitable API is then exposed using \mbox{OpenSHMEM} from within the
\mbox{OpenCL} kernel. The \mbox{OpenSHMEM} programming model is nested within
\mbox{OpenCL} and may be thought of as extending the latter. Developing
applications with this hybrid programming model will follow closely the
approach taken with \mbox{OpenCL}.
 
From an application development perspective, the \mbox{OpenCL} co-design model
is still used with no change in the development of \mbox{OpenCL} host code. It
is the \mbox{OpenCL} device programming API that is extended with
\mbox{OpenSHMEM}. In this way each \mbox{OpenCL} kernel would contain within it
a unique \mbox{OpenSHMEM} parallel job with a context inherited from the
\mbox{OpenCL} kernel. All initialization and allocation requirements in support
of the \mbox{OpenSHMEM} API are performed within the \mbox{OpenCL} kernel each
time it is enqueued for execution. Whereas \mbox{OpenCL} kernels are permitted
to communicate through global memory, no communication using the
\mbox{OpenSHMEM} API is permitted between kernels or between \mbox{OpenCL} work
groups. This follows from the \mbox{OpenCL} execution model that allows
synchronization within a work group but disallows synchronization between work
groups. The restriction upon synchronization between \mbox{OpenCL} work groups
has limited significance since the nested parallelism of \mbox{OpenCL} mode in
which work is distributed across multiple work groups containing multiple work
items can be ignored if a single work group is used. This simplification is
employed in the application of \mbox{OpenCL} to the Epiphany architecture.
Since the \mbox{OpenSHMEM} API is contained within the \mbox{OpenCL} device
kernel context, all \mbox{OpenSHMEM} memory allocation is only visible within a
kernel and is not persistent across multiple kernel invocations. This aspect of
the hybrid programming model could be revisited in the future but was
unnecessary for the initial demonstrations reported here.

It is worth addressing the issue of portability in the context of the proposed
hybrid programming model. As with the case of the use of non-standard
extensions originally employed to achieve good performance for \mbox{OpenCL}
development targeting Epiphany, the use of a hybrid \mbox{OpenCL} +
\mbox{OpenSHMEM} programming model will not be compliant with the \mbox{OpenCL}
standard and will not be portable to other architectures for which only a pure
\mbox{OpenCL} implementation exists. This issue cuts directly to the relevance
of standards in the development of high-performance code across differing
architectures. The very concept of performance-portability is questionable and
completely separate from that of portability in general. A code that is
non-standard and utilizes architecture-specific features is no less useful than
a code that is completely portable and compliant with a given programming
standard but achieves poor performance. For this reason, we contend that the
utility of programming standards such as \mbox{OpenCL} has less to do with
portability and more to do with providing programmers familiar syntax and
semantics for creating architecture-specific code. Therefore the lack of
general portability of our proposed programming model is not a significant
concern for programmers developing high-performance code.

\section{Application and Results}
\label{sec:application}

\begin{figure}
	\centering
		\includegraphics[width=1.0\textwidth]{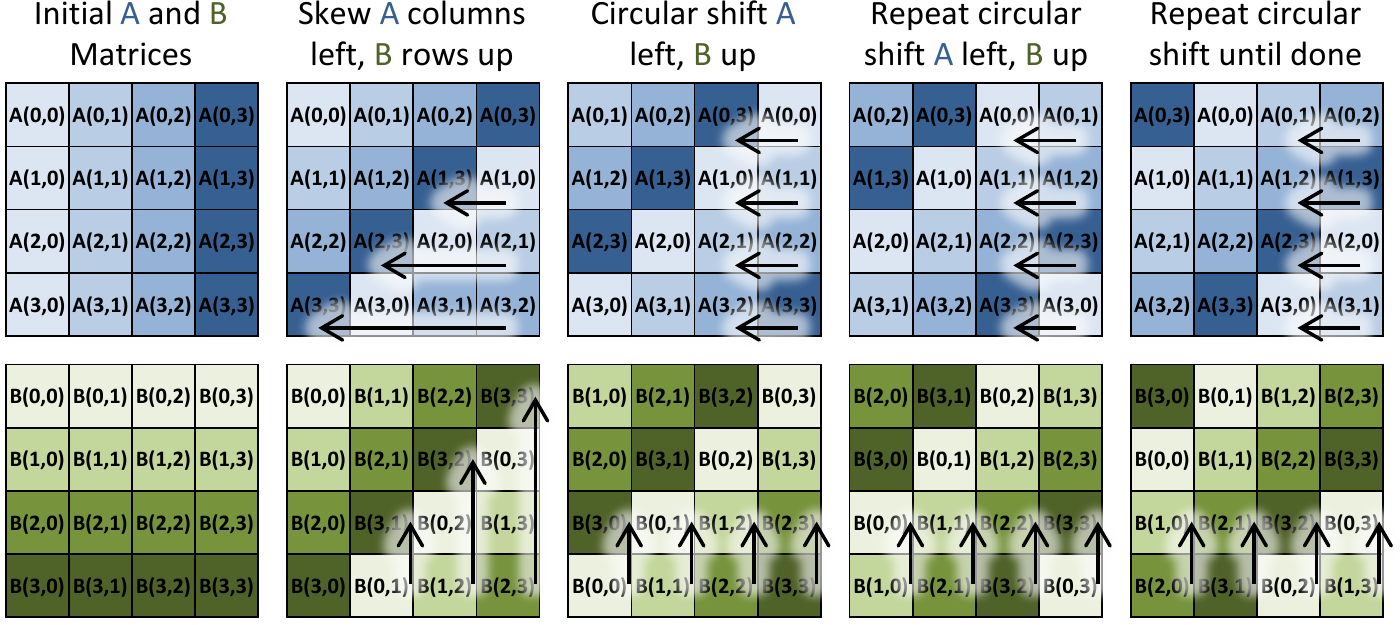}
	\caption{The 2D mesh network topology communication patterns for submatrix
skewing and shifting. A submatrix-submatrix multiplication occurs after each
communication step. For the Epiphany-III processor, this figure represents the
full inter-core communication pattern between the 16 cores on the device
although the communication pattern can be applied generally to larger or
smaller square arrays of cores. The initial skew communication may be
unnecessary if the submatrices are read in pre-skewed. An additional
communication step is needed to restore the shifted and skewed matrices if
desired, but this is unnecessary since a copy of the A and B matrices remains
within shared device memory.}
	\label{fig:cannon}
\end{figure}

Multiplication of matrices is a central building block in many scientific
applications. We apply the hybrid \mbox{OpenCL} + \mbox{OpenSHMEM} programming
model to matrix-matrix multiplication using the Cannon algorithm \cite{cannon}.
Cannon's algorithm exemplifies the use of 2D parallel decomposition to
effectively exploit this type of parallel architecture. The algorithm
decomposes a square matrix-matrix multiplication problem (C = A*B) across an
N-by-N collection of processing elements. Sub-matrices are shared between
neighboring processing elements after each submatrix-submatrix multiplication.
As illustrated in Figure \ref{fig:cannon}, the communication pattern begins by
skewing the columns of matrix A left and the rows of B upward within the 2D
mesh network topology.

For reference, a purely \mbox{OpenCL} implementation is benchmarked in which
each thread per core must read in submatrices from global memory. This
implementation lacks the data re-use that will lead to higher performance.
Instead of communicating submatrices for A and B to the left and upward,
respectively, equivalent bookkeeping is used to allow each thread to simply
read in the submatrix that is needed from global memory. The performance using
\mbox{OpenCL} alone achieves up to 794 MFLOPS for a matrix sizes of 128x128. It
is worth noting that the architecture is quite limited by the off-chip
bandwidth, particularly when loading memory directly rather than by using the
off-chip DMA engine (a feature not addressed by either \mbox{OpenCL} or
\mbox{OpenSHMEM} standards).

The same \mbox{OpenCL} code is then modified with \mbox{OpenSHMEM}. No changes
are required for the \mbox{OpenCL} host code. The \mbox{OpenSHMEM} header is
included in the \mbox{OpenCL} kernel, and the core-local buffers for matrices
A, B and C are allocated using \mbox{OpenSHMEM} semantics for symmetric shared
memory. The \mbox{OpenCL} kernel is further modified to use an \mbox{OpenSHMEM}
put call with appropriate barrier synchronization between threads to implement
the shifting of submatrices. The result is that a submatrix is read once from
global memory and then re-used. This is known to be necessary to achieve
optimal performance on the Epiphany architecture. The performance of the hybrid
\mbox{OpenCL} + \mbox{OpenSHMEM} programming model achieves up to 1812 MFLOPS.
With data re-use supported by \mbox{OpenSHMEM} the hybrid implementation easily
outperforms the reference \mbox{OpenCL}-only implementation. Performance for
this application is still limited by off-chip bandwidth, however, the inclusion
of the inter-core communication with the \mbox{OpenSHMEM} interface increases
performance by a factor of 2.3x. Results for various matrix sizes are shown in
Table \ref{tab:cannon}.

\begin{table}
	\centering
		\caption{On-chip matrix-matrix multiplication performance with pure
OpenCL and hybrid OpenCL + OpenSHMEM programming model}
		\label{tab:cannon}
		\begin{tabular}{l c c r}
			\hline
			\noalign{\smallskip}
\multirow{2}{*}{Matrix Size~~~~} & \multicolumn{2}{c}{Programming Model Performance (MFLOPS)} & \multirow{2}{*}{~~~~Speedup} \\ \cline{2-3}
			\noalign{\smallskip}
 & OpenCL & OpenCL + OpenSHMEM & \\
			\hline
			\noalign{\smallskip}
  $32\times32$ & 218 &  504 & 2.3x \\
  $64\times64$ & 424 & 1000 & 2.4x \\
$128\times128$ & 794 & 1817 & 2.3x \\
			\hline
		\end{tabular}
\end{table}

\section{Conclusions and Future Work}
\label{sec:conclusions}

We have proposed and demonstrated a hybrid \mbox{OpenCL} + \mbox{OpenSHMEM}
programming model for device-level parallel programming architectures like the
low-power Epiphany RISC array processor. This hybrid model directly resolves
the most critical deficiency encountered in the use of \mbox{OpenCL} alone for
this architecture. The introduction of \mbox{OpenSHMEM} allows the proper
management of the on-chip distributed symmetric shared memory, which is
critical for obtaining high performance with this architecture. Benchmarks for
matrix-matrix multiplication demonstrate that the hybrid programming model can
achieve better performance for this architecture and substantially outperforms
the use of \mbox{OpenCL} alone.

\bibliography{OpenFrankenstein}

\end{document}